**Controlling the biodegradation rates of poly(globalide-co-ε-caprolactone) copolymers by post polymerization modification**

Camila Guindani [a,b], Graziâni Candiotto [c,d], Pedro H. H. Araújo [a], Sandra R. S. Ferreira [a], Débora de Oliveira [a], Frederik R. Wurm [b,*], Katharina Landfester [b,*]

[a] Department of Chemical Engineering and Food Engineering - Federal University of Santa Catarina - EQA/UFSC - C.P. 476, CEP 88040-900, Florianópolis, SC, Brazil.

[b] Max Planck Institute for Polymer Research, Ackermannweg 10, 55128 Mainz, Germany

[c] Department of Physics – Federal University of Paraná – C.P. 19044, CEP 81531-980, Curitiba, PR, Brazil

[d] Institute of Chemistry – Federal University of Rio de Janeiro, CEP 21941-909, Rio de Janeiro, RJ, Brazil

*Corresponding author: wurm@mpip-mainz.mpg.de

**ABSTRACT**

Controlling the degradation rates of polymers is crucial for their application in tissue engineering or to achieve degradation of the polymers in the wastewater purification. As hydrophobic polyesters often exhibit very slow degradation rates, we report here increased biodegradation rates of poly(globalide-*co*-ε-caprolactone) copolymers (PGlCL) produced by enzymatic ring-opening copolymerization and post-functionalized with *N*-acetylcysteine by thiol-ene reaction. The degradation rates of the PGlCL and post-modified PGlCL-NAC films were determined by weight-loss experiments. The polymer films were immersed in phosphate-buffered saline (PBS) solution, and PBS containing lipase from *Pseudomonas cepacia*. The degree of functionalization affected the degradation behavior, and samples with a higher degree of functionalization presented higher weight loss. Finally, a degradation assay was performed in activated sludge, and PGlCL-NAC presented high degradability, having a degradation behavior similar to starch. Density Functional Theory (DFT) calculations were used to assess the changes in chemical properties and electronic charge distribution of PGlCL after its functionalization with NAC, helping to understand its influence in their degradability. The results obtained confirm the possibility to increase the degradation rates of copolyesters based on caprolactone and globalide by thiol-ene post-functionalization, being a promising alternative for applications in biomedicine or the packaging sector.

# 1. INTRODUCTION

Degradable polymers are important as biomedical polymers, contributing to treatments, in which the material needs to be resorbed over time, such as in tissue regeneration and drug delivery applications; biodegradable polymers can also play an important role to decrease plastic waste pollution [1].

Polycaprolactone (PCL) is an example of a biodegradable polyester that can be used in both biomedical and polymer packaging applications, thanks to its excellent properties such as high flexibility, ease of molding, degradability, and biocompatibility, as well as low toxicity [2,3]. However, its long-term degradation (2-4 years) under physiological conditions is considered a disadvantage, especially for biomedical applications, limiting its potential applications [2]. PCL is degraded by the hydrolysis of the ester linkages, but its high degree of crystallinity and hydrophobicity reduces water uptake into the polymer matrix and slows down the degradation process [2,4]. Polyglobalide (PGl) is another example of a polyester, which is considered as a biocompatible polymer with excellent mechanical properties [5–7]. The presence of olefins in the polymer backbone enables the post-polymerization functionalization, e.g. by thiol-ene coupling reaction. Copolymers based on ε-caprolactone and globalide allow the preparation of multifunctional polyesters, which allow further adjusting of materials properties depending on the target application.

In this work, poly(globalide-*co*-ε-caprolactone) (PGlCL) copolymers were produced by enzymatic ring-opening polymerization (e-ROP) in supercritical carbon dioxide (scCO$_2$) with two different monomer ratios. Subsequently, the globalide units of PGlCL were post-modified with *N*-acetylcysteine (NAC) via thiol-ene coupling, aiming to accelerate the degradation times. The covalent conjugation of NAC to PGlCL (PGlCL-NAC) by thio-ether linkages was previously studied [8] and has proven to reduce hydrophobicity and the degree of crystallinity of the final material. Besides, the presence of thio-ether groups confers an antioxidant potential to PGlCL-NAC, which is an interesting characteristic for applications in biomedicine and in the plastic packaging sector.

The hydrolytic, enzymatic, and microbial degradation of PGlCL and PGlCL-NAC copolymers were evaluated (Figure 1). Additional theoretical studies provided further understanding of the stability and degradability of PGlCL and post-modified PGlCL [9],[10]. To the best of our knowledge, these are the first studies on the degradation of PGlCL and modified PGlCL evaluated in different systems. We believe that the results reported in this work are an important contribution to the development of cleaner chemical processes and readily degradable materials.

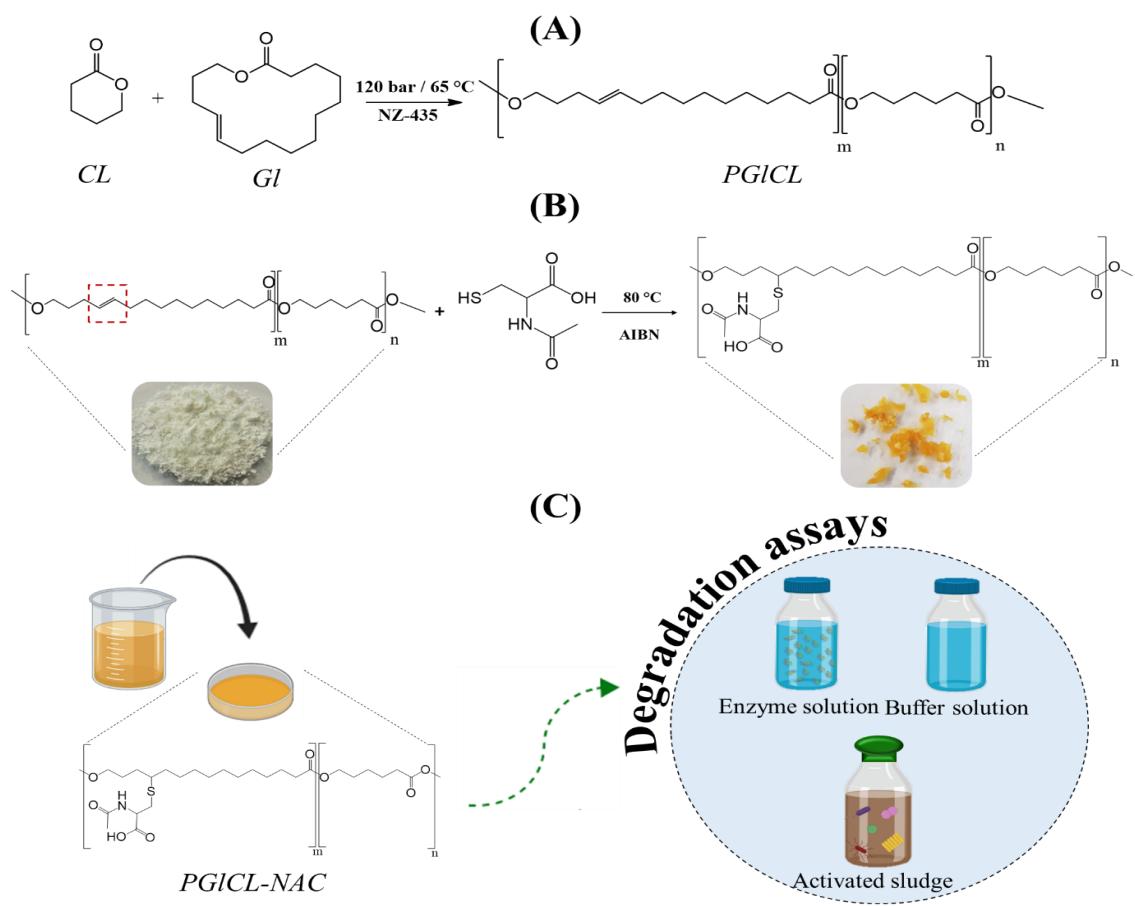

Figure 1. (A) Enzimatic ring-opening copolymerization of ε-caprolactone and globalide in supercritical $CO_2$ (B) Poly(globalide-*co*-ε-caprolactone) side-chain functionalization with *N*-acetylcysteine by thiol-ene reaction. (C) Assessment of the degradation of PGlCL-NAC in different media.

## 2. MATERIAL AND METHODS

### 2.1. Materials

Dichloromethane P.A. 99.8% (DCM), ethanol P.A. 99.8% (EtOH), chloroform P.A. 99.8%, and glacial acetic acid P.A. 99.8% were purchased from Merck (Germany). The free radical initiator azobisisobutyronitrile 98% (AIBN) was purchased from Sigma Aldrich (Germany). Carbon dioxide (99.9% purity) used as solvent was purchased from White Martins A/S, Brazil. N-acetylcysteine 99.8% (NAC), phosphate-buffered solution (PBS) (0.1 M, pH 7.4), and the commercial lipase from *Pseudomonas cepacia* were purchased from Sigma Aldrich (Germany) (enzymatic activity $\geq 30$ U g$^{-1}$, provided by the manufacturer). Novozym 435 (NZ-435) was kindly donated by Novozymes, Brazil, A/S (commercial lipase B from *Candida antarctica* immobilized on cross-linked polyacrylate beads, esterification activity 42 U/g, measured according to a procedure adapted from literature [11]). Enzymes were dried under vacuum (0.4 bar) and 70 °C, during 16 h [12] and stored in a desiccator over silica and 4 Å molecular sieves. Globalide (Gl) was a kind gift of Symrise. ε-caprolactone was purchased from Sigma-Aldrich (Germany). Both globalide and ε-caprolactone (CL) were dried under vacuum (0.1 bar) and 100 °C, during 24 h [12] and also stored in a desiccator over silica and 4 Å molecular sieves.

### 2.2. Poly(globalide-*co*-ε-caprolactone) synthesis using supercritical carbon dioxide

Polymerization experiments were carried out as previously described by Guindani et al. [13] using supercritical carbon dioxide (scCO$_2$) as a solvent. The pressure and temperature of the system were maintained constant at 120 bar and 65 °C and the reaction was carried out for 2 h. Enzyme content was fixed at 5 wt% (relative to the total monomer amount), and the CO$_2$ to monomers mass ratio was fixed at 1:2. Two different globalide/ε-caprolactone monomer ratios were tested: PGlCL 10/90 and PGlCL 25/75. After polymerization, the material was purified through solubilization in DCM, followed by the separation of the enzymes and precipitation of the polymer in cold EtOH. DCM and EtOH were used at the volumetric proportion of 1:6. The polymeric suspension was filtered and dried at room temperature *in vacuo* to constant weight.

**2.3. Thiol-ene functionalization of poly(globalide-*co*-ε-caprolactone) with N-acetylcysteine**

PGlCL copolymers produced in $scCO_2$, with different Gl/CL ratios (10/90 and 25/75) were functionalized with N-acetylcysteine (NAC) by thiol-ene reactions. The reactions were carried out in an oil bath at 80 °C, for 24 h, under continuous magnetic stirring and nitrogen atmosphere. The amount of NAC used was established as being twice the minimum amount of NAC required to react with all double bonds. AIBN content was fixed in 5% (mol), relative to the NAC amount. A mixture of acetic acid and chloroform in a volumetric proportion of 3:1 (acetic acid: chloroform) was used as a solvent, and the polymer concentration was kept constant at 75 mg/mL.

After polymerization, the solvents were removed in an air circulation oven. The dried material was washed in cold water, for the free NAC removal, followed by a second wash in methanol for AIBN removal. PGlCL modified with NAC (PGlCL-NAC) was then dried under vacuum at room temperature, up to constant weight. PGlCL double bond consumption was determined in previous work by Guindani and collaborators [8] through proton nuclear magnetic resonance spectroscopy ($^1$H NMR), by comparing the integral values of the peaks related to the double bonds before and after the thiol-ene reaction.

**2.4. Preparation of PGlCL and PGlCL-NAC films**

PGlCL and PGlCL-NAC films were prepared by casting 120 µL of a polymeric solution (120 mg/mL) onto microscope coverslips with a 25 mm diameter. PGlCL samples were diluted in chloroform, while PGlCL-NAC was diluted in a chloroform:ethanol mixture 9:1 (v/v). The samples were dried under vacuum at room temperature until a constant weight was achieved. The weights of the films were ca. 11-15 mg.

**2.5. Contact angle assay**

PGlCL and PGlCL-NAC films were evaluated regarding its hydrophilicity/hydrophobicity. The contact angle between the polymeric films and water droplets were measured in a goniometer (Ramé-Hart Instrument Co. - Ramé-Hart 250). All measurements were performed in triplicate at room temperature with a drop volume of 10 µL, after a stabilization time of 3 s.

**2.6. Degradation of PGlCL and PGlCL-NAC films in PBS and lipase solutions**

The degradation experiments were conducted at 37 °C by immersing each polymer film with the coverslips in 4 mL PBS solution (0.1 M, pH 7.4). In the case of degradation by lipase, the films were immersed in PBS solution containing 0.05 mg lipase from *Pseudomonas cepacia*/mL PBS solution. The PBS-enzymatic solution was changed every 24 h to maintain the enzymatic activity. The coverslips with the polymer films were picked up over predetermined time intervals, washed with distilled water and dried under vacuum. The weights of the microscope coverslips were measured before film casting and the weight of the coverslips with polymer was measured before and after degradation studies.

**2.7. Morphology of PGlCL and PGlCL-NAC films - Scanning electron microscopy (SEM)**

The morphology of the surface of the film before and after degradation was observed with a Gemini 1530 (Carl Zeiss AG, Oberkochem, Germany) scanning electron microscope operating at an accelerating voltage of 0.121 kV.

**2.8. Gel permeation chromatography (GPC)**

GPC measurements were performed in THF with a PSS SecCurity system (Agilent Technologies 1260 Infinity). Sample injection was performed using a 1260-ALS autosampler (Waters) at 30 °C. SDV columns (PSS) with dimensions of 300 × 80 mm, 10 μm particle size, and pore sizes of 106, 104, and 500 Å were employed. A DRI Shodex RI-101 detector (ERC) and UV−vis 1260-VWD detector (Agilent) were used for detection. Calibration was achieved using polystyrene standards provided by Polymer Standards Service.

**2.9. Biodegradability in activated sludge**

In this assay, polymer biodegradability in activated sludge was assessed using the method based on the Organization for Economic Cooperation and Development (OECD) guideline 301F [14]. In summary, the biological oxygen demand (BOD) for each chemical was measured using the OxiTop control manometric closed system (WTW, Germany) over 28 d. The percentage of biodegradability was determined by comparing the measured BOD and the calculated theoretical oxygen demand (ThOD) values. In a typical run, flasks containing the following compositions were used: (1) Two flasks for the inoculum "blank", containing activated sludge and mineral medium with nutrients; (2) Two flasks for procedure control, containing the activated sludge, a reference compound readily biodegradable (in this case it was used starch) and mineral medium; (3) One flask for toxicity control, containing the inoculum, PGlCL-NAC 25/75, starch and mineral medium; (4) One flask for evaluating the biodegradability of the tested material, containing activated sludge, PGlCL-NAC 25/75 and mineral medium.

The activated sludge sample used in all studies was from the same batch of collected at the waste treatment unit in the city of Mainz (Mombach), Germany and it was aerated in the dark at 20 °C for 7 d before the start of the experiments. Activated sludge (2.8 mL) with a solid content of 3.9 g/L was mixed to 365 mL of mineral medium in amber bottles, giving a final concentration of approximately 30 mg/L, as described by the OECD procedure. It was used 24.16 mg of starch and 13 mg of PGlCL-NAC 25/75 per bottle. Magnetic stirrer bars were also added. The screw-top measuring heads, containing sodium hydroxide pellets to adsorb produced carbon dioxide, were replaced. The flasks were stirred in an incubator cabinet in the dark at 20 °C. Oxygen consumption data via measurement of pressure loss were recorded over 28 d. ThOD of each sample was calculated as described in the OECD 301 guidelines [14]. Blank oxygen consumption values (BOD values representing background respiration in activated sludge) were deducted from the BOD of the test compound before determining the percentage biodegradability (see Support Information, Equations S1, S2, and S3).

## 2.10. DFT calculations

In this work, all three-dimensional molecular structures of the copolymers PGlCl and PGlCl-NAC were generated and analyzed by Avogadro [15] (version 1.2.0). The theoretical results for the copolymers were performed into two steps. In the first step, it was performed the calculations related to the geometry optimization of PGlCL and PGlCL-NAC structures and determination of properties such as the dipole moments, Gibbs free energy ($G$) and partition coefficient ($P$). To determine Gibbs free energy for PGlCL and PGlCL-NAC for partition coefficient calculation, the molecular structures of the polymers were optimized in the gas-phase, and also in water and n-octanol. The optimized structures were confirmed as real minima by vibration analysis (no imaginary frequency was detected). The results were obtained at normal temperature and pressure (NTP) using B3LYP hybrid functional of DFT in conjunction with 6-311++G** basis set, using water and n-octanol as solvent through the SMD model.

The second step was focused on assessing the electronic charge distribution and global chemical reactivity descriptors for PGlCL and PGlCL-NAC. The results for this step were obtained in gas-phase at normal temperature and pressure (NTP) using B3LYP hybrid functional of DFT in conjunction with 6-311++G** basis set. All DFT calculations were performed using the gaussian 09 package program.

## 3. RESULTS AND DISCUSSION

### 3.1. Degradation study in PBS and enzymatic solutions

PGlCL and PGlCL-NAC films were prepared by drop-casting and subjected to degradation assays. The characterization of the materials regarding their affinity for water were carried out through contact angle assays. Other characterization assays of PGlCL and PGlCL-NAC samples were reported in our previous studies [8,13].

*3.1.1. Weight loss data assessment*

For degradation assays, the films were immersed in PBS solution (0.1 M, pH 7.4) and enzymatic solution (lipase from *Pseudomonas cepacia*, 0.05 mg/mL in PBS), and the weight loss of the films were assessed over several days. Figure 2 presents the degradation curves and contact angle with water for (A) PGlCL 10/90 and PGlCL-NAC 10/90 and for (B) PGlCL 25/75 and PGlCL-NAC 25/75.

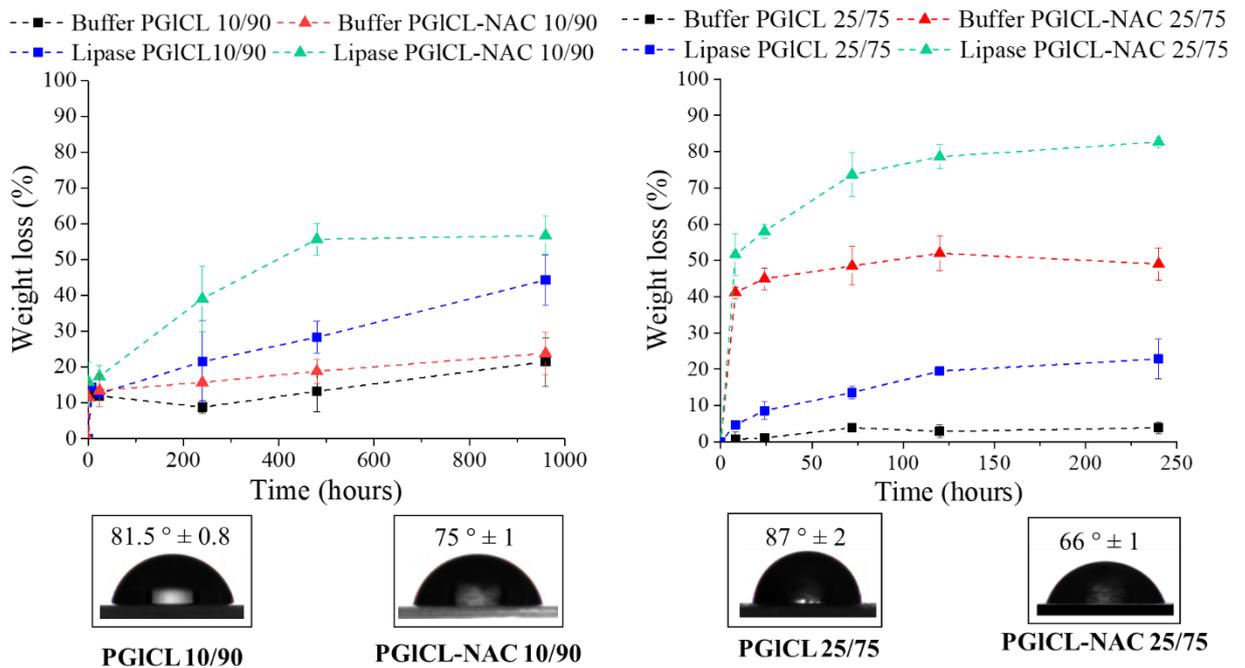

Figure 2. (A) Contact angle and degradation curves of PGlCL 10/90 and PGlCL-NAC 10/90 in PBS and in lipase solution; (B) Contact angle and degradation curves of PGlCL 25/75 PGlCL-NAC 25/75 in PBS and in lipase solution.

The contact angle images indicate a variable hydrophobicity of all samples. While PGlCL 25/75 proved to be more hydrophobic as PGlCL 10/90 due to the higher amount of globalide units, the post-modified polymer exhibited increased hydrophilicity due to the relatively hydrophilic NAC moieties, which influence the degradation behavior (hydrolysis and solubilization of the chains). All polyesters proved a significant degradation over the investigated period. The incorporation of more globalide units (or NAC-modified globalides) increased the degradation both in the buffer and in the presence of enzymes. Looking at each sample individually, degradation in the lipase solution resulted in higher degradation rates, in comparison to the degradation carried out in PBS solution. *Pseudomonas cepacia* lipase can catalyze the cleavage of ester bonds [16], resulting in faster degradation and higher weight loss of the polymer, in comparison to the degradation in PBS. When comparing the degradation behavior of PGlCL 10/90 and PGlCL-NAC 10/90 in PBS solution, the weight loss profile for both samples was similar, reaching around 20% weight loss after 40 d (960 h.) However, for degradation in the lipase solution, the functionalization of PGlCL 10/90 with NAC increased the weight loss rate, in comparison to non-functionalized PGlCL 10/90. After 20 d (480 h) of degradation, PGlCL-NAC 10/90 reached around 55% of weight loss, while for PGlCL 10/90 non-functionalized the weight loss was close to 28%.

For PGlCL 25/75 (Figure 2B), the effect in the weight loss caused by modification with NAC was even more evident. In PBS, PGlCL-NAC 25/75 reached around 50% weight loss after 10 d, while for non-functionalized PGlCL 25/75, only around 5% was reached after 10 d. In comparison to the degradation profile of PGlCL 10/90 (Figure 2A), PGlCL 25/75 (Figure 2B) degraded slower, which can be related to the higher content of globalide units, which are more hydrophobic. PGlCL 25/75 presented a water contact angle value of 87° ± 2, while for PGlCL 10/90 films, a contact angle of 81.5° ± 0.8 was determined. Besides, PGlCL 25/75 exhibited a higher degree of crystallinity compared to PGlCL 10/90, [13] it also influences the degradation behavior of the polyesters. The hydrophilicity strongly affects the water uptake, which means that hydrophobic polymers have limited water uptake and degrade slower [17,18]. The presence of crystalline domains reduces the ability of the water to penetrate into the polymer matrix, decreasing degradation rates [18,19].

For enzymatic degradation, after 240 h PGlCL 25/75 reached around 20% weight loss, while PGlCL-NAC 25/75 reached more than 80%. Vidaurre et al. [20] reported 80% weight loss for the enzymatic degradation of poly(ε-caprolactone) films in lipase from *Pseudomonas fluorescens* after 240 h but using a 2-fold higher lipase concentration, which emphasizes the ready enzymatic degradation of our PGlCL-NAC. According to Guindani et al. [8], with increasing number of NAC-modified globalides also the crystallinity of the copolymers decreased and the overall hydrophilicity increased, which was detected by decreased water contact angles (PGlCL-NAC 10/90 = 75° ± 1 (Figure 2A) and PGlCL-NAC 25/75 = 66° ± 1 (Figure 2B)).

*3.1.2. Visualization of PGlCL and PGlCL-NAC (Gl/CL = 25/75) degraded films by scanning electronic microscopy (SEM)*

Besides weight loss profiles, also morphological changes during the degradation process were visualized by scanning electron microscopy (SEM) (PGlCL 25/75 (Figure 3A) and PGlCL-NAC 25/75 (Figure 3B) films before and after degradation).

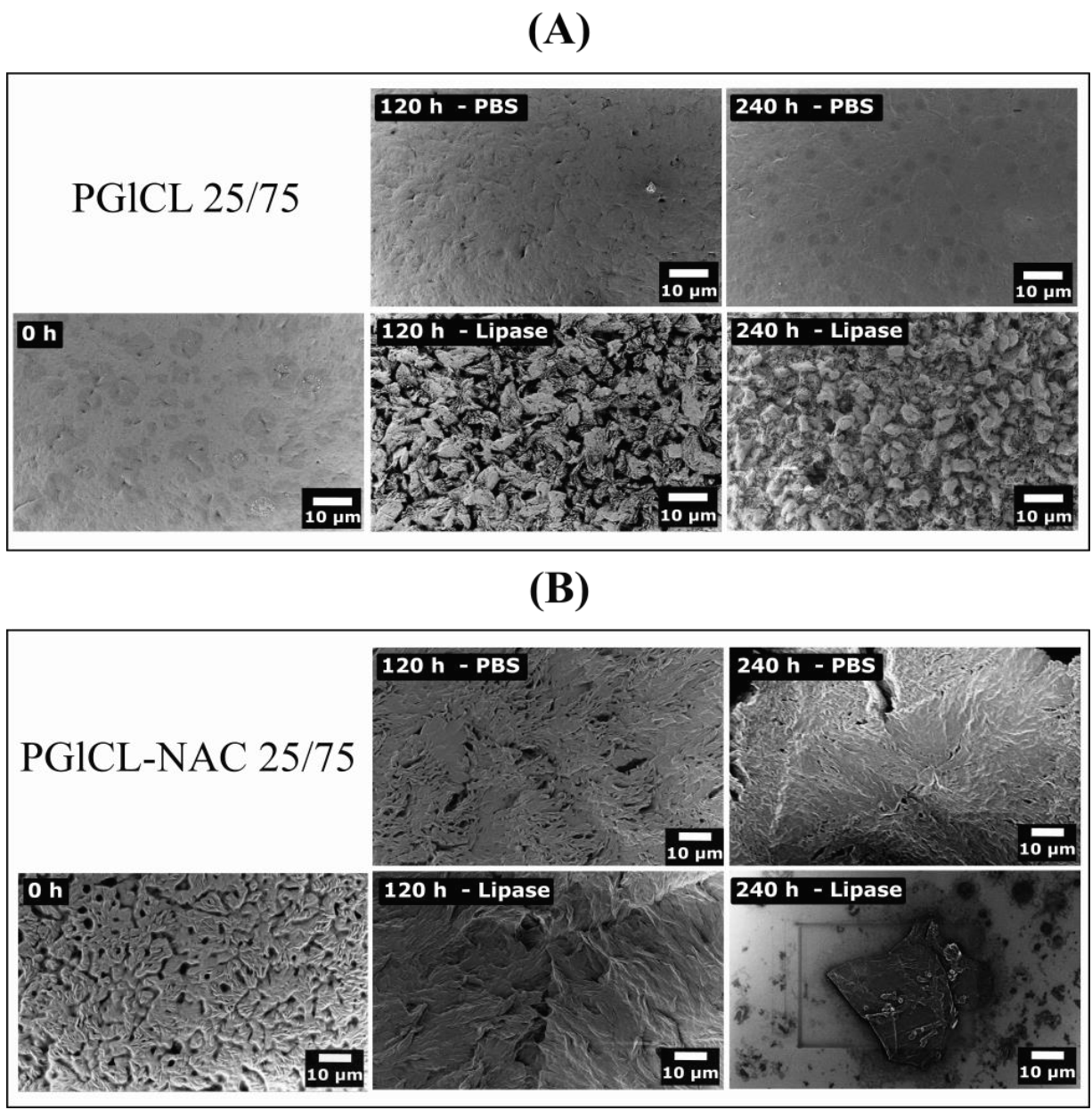

Figure 3. Scanning electronic microscopy images of (A) PGlCL and (B) PGlCL-NAC in a Gl/CL ratio of 25/75 before and after degradation in lipase and PBS solutions.

The morphological changes and the appearance of small cracks and holes indicate a surface erosion during the degradation period for both PGlCL 25/75 and PGlCL-NAC 25/75. When treated with lipase, the film's surface proved higher decomposition compared to PBS treatments (Figure 3A), which was in accordance with the weight-loss experiments. Vidaurre et al. [20] and Montaudo and Rizzarelli [21] also observed an increased roughness of the surface of polyester films during enzymatic degradation. It is important to mention that the initial roughness of the PGlCL-NAC films was higher compared to the unmodified polymers, which might be rationalized by the additional polar NAC-groups and also be a reason for increased degradation rates, which are visible in the SEM images for lipase and PBS treatment (Figure 3B) [22]. The SEM images underline that PGlCL-NAC presented a higher degradation rate than non-functionalized PGlCL, probably due to the higher hydrophilicity and increased surface roughness of the polymer films.

*3.1.3. Molecular weights*

Molecular weight of PGlCL and PGlCL-NAC was evaluated before degradation (t = 0 h) and after degradation in buffer (t = 960 h) and lipase (t = 960 h) (Figure 4). The NAC-modification does not change the apparent molecular weight determined in GPC. Both materials, before and after modification, proved a significant decrease of the molecular weight when treated with the aqueous buffer or the enzyme. Together with the weight loss curves and the SEM images (Figures 2 and 3), it is clear that there are two factors involved on the degradation of the polymeric films [23,24]: (1) the solubilization of the lower molecular weight chains in aqueous media (decrease of the dispersity) and (2) the hydrolysis of the polymer chains (decrease of the molecular weight). In a polymer degradation process, the weight loss occurs when the chains or chain fragments (originally present in the sample, or formed after hydrolysis) are solubilized by the aqueous media [24,25]. In this case, the functionalization of PGlCL with NAC helps improving the degradation of the polymer not only because of easing the access of water among the polymer chains and promoting faster hydrolysis, but also because of conferring to the material some solubility in aqueous media.

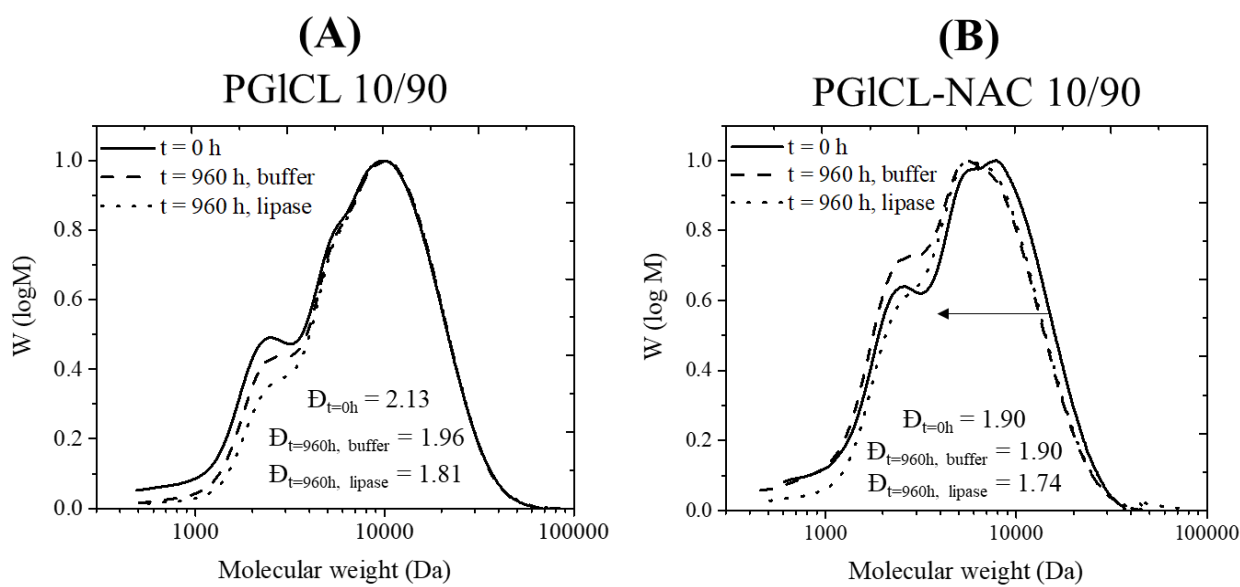

Figure 4. Molecular weight distribution (MWD) curves of (A) PGlCL 10/90, (B) PGlCL-NAC 10/90 before and after degradation in lipase and PBS solutions. Straight line: MDW of the sample before degradation; dashed line: MWD of the sample after degradation in buffer; dotted line: MWD of the sample after degradation in lipase solution.

**3.2. Degradation assay in activated sludge**

To evaluate the biodegradation of PGlCL-NAC 25/75 during the wastewater treatment, we studied the biodegradation in activated sludge provided from the local sewage plant in Mainz (Germany). A manometric respirometry assay was performed and the oxygen consumption was determined, as well as the percentage of degradation. Oxygen consumption raw data are presented in the Support Information (Table S1); Figure 5 summarizes the polymer degradation process by microorganisms schematically and plots the degradation curves of PGlCL-NAC 25/75, a mixture of PGlCL-NAC 25/75 and starch as a toxicity control, and pure starch as a positive control. Starch as a readily biodegradable biopolymer was used as a control and was degraded after 28 d to ca. 75%. The mixture PGlCL-NAC 25/75 + starch was used as a toxicity control, and the results clearly show that PGlCL-NAC 25/75 did not present toxicity to the microorganisms in the activated sludge.

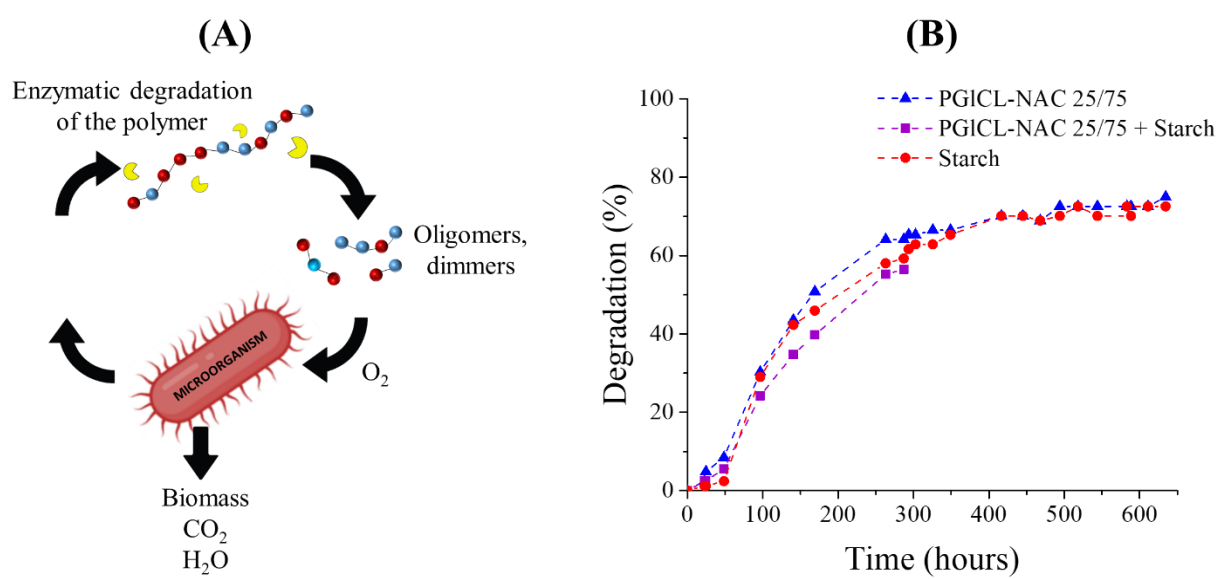

Figure 5. (A) Theoretical process of polymer degradation through the action of microorganisms present in activated sludge (B) Degradation curves obtained by the OECD biodegradability assays in activated sludge. Degradation is expressed as the percentage of the oxygen amount taken up by the microbial population during biodegradation (BOD), in relation to the total amount of oxygen required to oxidize the test substance (ThOD) (Equation S3, SI). PGlCL-NAC 25/75 (blue triangles), PGlCL-NAC 25/75 + Starch (purple squares), Starch (red circles), and the blank solution (black squares).

OECD guidelines state that the test substance will be considered inhibitory if the degradation of the systems containing both the test and positive control substance, reaches less than 25% by day 14 [14]. As Figure 5B shows, the toxicity control system reached 25% degradation on day 4, which means PGlCL-NAC 25/75 is not considered toxic in the concentration used in the test. PGlCL-NAC 25/75 presented a degradation curve very similar to starch, also reaching almost 75% of degradation. These results show that PGlCL-NAC is a highly biodegradable material and have a potential for biomedical and other applications, in which readily biodegradable polymers are demanded.

### 3.3. Theoretical considerations

DFT calculations were performed to estimate important properties of PGlCL and PGlCL-NAC, such as dipole moment, Gibbs free energy of solvation and partition coefficient, which helps to understand what is the influence of the NAC-modification on the degradation kinetics of the copolymers. Besides, DFT calculations also allowed the assessment of information about the electronic charge distribution of these structures. This enables us to investigate at the atomic level how the presence of NAC affects the chemical reactivity of the copolymer, and which chain sites are more susceptible to hydrolysis [26].

The theoretical geometry optimization was performed by DFT for PGlCL and post-modified PGlCL-NAC is summarized in the Supporting Information. To simplify the simulation process, two repeat units were analyzed, based on one ring-opened globalide unit (functionalized with NAC for PGlCL-NAC) and one ε-caprolactone unit. This simplification should provide a good approximation to understand how the attachment of NAC alters the degradation behavior. Molecular symmetry can be used to predict many molecular properties such as its dipole moment, atomic charge, bond length, and dihedral angles. The functionalization of PGlCL with NAC increases the dipole moment (see Table 1) and thus increases the hydrophilicity.

To quantify the increase in hydrophilicity, the partition coefficient values (represented in the logarithm form, $log\ P$) of the copolymers and the NAC-modified copolymers were calculated based on the dimer approximation. The partition coefficient ($P$) is defined as the ratio between the concentration of a solute in two phases of a mixture of two immiscible solvents at equilibrium [27], and it is dependent on the Gibbs free energy of solvation. To determine $log\ P$ values, the Gibbs free energy ($G$) and the Gibbs free energy of solvation ($\Delta G_{solv}$) were first calculated by DFT simulation. The theoretical prediction of partition coefficient logarithm for n-octanol/water mixture ($log\ P^{O/W}$) were calculated according to Equation (1) [28].

$$log\ P^{O/W} = \frac{\Delta G_{solv}^{W} - \Delta G_{solv}^{O}}{2.303 RT} \tag{1}$$

In this equation, $\Delta G_{solv}$ is the Gibbs free energy of solvation for two different phases, where the superscript labels $W$ and $O$ are respectively the water and n-octanol solvents, $R$ is the gas constant (8.314 J/K⁻mol⁻¹) and $T$ is the room temperature (298.150 K). The knowledge of these properties provides information about the hydrophilicity [27], which is a determining factor for understanding the degradation behavior of PGlCL and PGlCL-NAC. The calculated values for Gibbs free energy and $log\ P^{O/W}$ are shown in Table 1. The functionalization of PGlCL with NAC causes a decrease in $log\ P^{O/W}$ values, i.e. an increase in hydrophilicity and thus a higher water uptake.

Calculations related to frontier molecular orbitals (FMOs) and electronic charge distribution of PGlCL and PGlCL-NAC were also performed to understand, which ester bonds are more susceptible to hydrolysis (Details in section 2.2 of the SI). According to the calculations, after functionalization with NAC, PGlCL's energy gap between FMOs ($E_{GAP} = E_{Lumo} - E_{Homo}$, where Lumo = lowest unoccupied molecular orbital and Homo = highest occupied molecular orbital) decreased in comparison to non-functionalized PGlCL (Table S6). Compounds with low gap energy are chemically reactive as electrophiles and tend to be subject to nucleophilic substitution [29,30], which means hydrolysis is more likely to occur for PGlCL-NAC when compared to PGlCL. Additionally, the charge distribution proves that after functionalization with NAC, PGlCL presented an increase in the positive potential distribution along the molecular chain (Figure S3 and S4), getting more susceptible to nucleophilic attack, especially in the carbonyl group. This means that ester linkages located close to NAC structure are more likely to be hydrolyzed compared to other ester linkages in the copolymer [31].

## 4. CONCLUSION

The post-polymerization modification of unsaturated polyester copolymers (poly(globalide-*co*-ε-caprolactone)) with *N*-acetylcysteine by thiol-ene reaction increased the biodegradation rates.

The degradation profiles of PGlCL and PGlCL-NAC films in PBS and lipase solutions were determined, proving that the modification of PGlCL with NAC increased the degradability of the materials. PGlCL-NAC were also subjected to degradation in activated sludge proving them to be readily biodegradable under these conditions. These increased degradation rates can be explained by increased hydrophilicity after the modification with NAC, which was confirmed by water contact angles and additional calculations, e.g. of $log\ P$ values or charge-distributions, which allowed also to estimate the place of chain hydrolysis. The data presented in this work underline that post-polymerization modifications of unsaturated polyesters by thiol-ene reactions are a simple and feasible strategy that might be used to accelerate degradation rates, e.g. for biomedical applications or to ensure degradation during wastewater treatment.

**Acknowledgments**

The authors thank Symrise (Brazil) for kindly supply the monomer globalide, and Gunnar Glaßer (MPIP) for the SEM images. We gratefully acknowledge CAPES (Coordenação de Aperfeiçoamento de Pessoal de Nível Superior) / Programa Doutorado CAPES-DAAD-CNPQ /Processo nº 88887.161406/2017-00 for the scholarship to C. Guindani, and CAPES-PRINT, Project number 88887.310560/2018-00 for the financial support. F.R.Wurm thanks the German Federal Ministry for Education and Research (BMBF) for the support of the program "Research for sustainable development (FONA)", "PlastX – Plastics as a systemic risk for social-ecological supply systems" (grant number: 01UU1603A).

**REFERENCES**

[1]  Y. Ikada, H. Tsuji, Biodegradable polyesters for medical and ecological applications, Macromol. Rapid Commun. 21 (2000) 117–132.

[2]  M.A. Woodruff, D.W. Hutmacher, The return of a forgotten polymer - Polycaprolactone in the 21st century, Prog. Polym. Sci. 35 (2010) 1217–1256. doi:10.1016/j.progpolymsci.2010.04.002.

[3]  A.C. Albertsson, R.K. Srivastava, Recent developments in enzyme-catalyzed ring-opening polymerization, Adv. Drug Deliv. Rev. 60 (2008) 1077–1093. doi:10.1016/j.addr.2008.02.007.

[4]  H. Jeon, H. Lee, G. Kim, A Surface-Modified Poly(ε-caprolactone) Scaffold Comprising Variable Nanosized Surface-Roughness Using a Plasma Treatment, Tissue Eng. Part C Methods. 20 (2014) 951–963. doi:10.1089/ten.tec.2013.0701.

[5]  M.L. Focarete, M. Scandola, A. Kumar, R. a Gross, Physical Characterization of Poly (ω-pentadecalactone ) Synthesized by Lipase-Catalyzed Ring-Opening Polymerization, J. Polym. Sci. Part B Polym. Phys. 39 (2001) 1721–1729. doi:10.1002/polb.1145.

[6]  M. de Geus, I. van der Meulen, B. Goderis, K. van Hecke, M. Dorschu, H. van der Werff, C.E. Koning, A. Heise, Performance polymers from renewable monomers: high molecular weight poly(pentadecalactone) for fiber applications, Polym. Chem. 1 (2010) 525. doi:10.1039/b9py00360f.

[7]  J. Cai, C. Liu, M. Cai, J. Zhu, F. Zuo, B.S. Hsiao, R.A. Gross, Effects of molecular weight on poly(ω-pentadecalactone) mechanical and thermal properties, Polymer (Guildf). 51 (2010) 1088–1099. doi:10.1016/j.polymer.2010.01.007.

[8]  C. Guindani, P. Dozoretz, P.H.H. Araújo, S.R.S. Ferreira, D. de Oliveira, N-acetylcysteine side-chain functionalization of poly(globalide-co-ε-caprolactone) through thiol-ene reaction, Mater. Sci. Eng. C. 94 (2019).

doi:10.1016/j.msec.2018.09.060.

[9] P. Hohenberg, W. Kohn, Inhomogeneous Electron Gas, Phys. Rev. 136 (1964) B864–B871. doi:10.1103/PhysRev.136.B864.

[10] Y. Zhu, D. Zhang, Z. Zhang, Z. Wang, The Effect of Polymer Structures on Complete Degradation: A First-Principles Study, ChemistryOpen. 7 (2018) 463–466. doi:10.1002/open.201800078.

[11] D. Oliveira, A.C. Feihrmann, A.F. Rubira, M.H. Kunita, C. Dariva, J.V. Oliveira, Assessment of two immobilized lipases activity treated in compressed fluids, J. Supercrit. Fluids. 38 (2006) 373–382. doi:10.1016/j.supflu.2005.12.007.

[12] S.R. Comim Rosso, J.G. Veneral, D. de Oliveira, S.R.S. Ferreira, J.V. Oliveira, Enzymatic synthesis of poly(ε-caprolactone) in liquified petroleum gas and carbon dioxide, J. Supercrit. Fluids. 96 (2015) 334–348. doi:10.1016/j.supflu.2014.07.004.

[13] C. Guindani, P. Dozoretz, J.G. Veneral, D.M. Silva, P.H.H. Araújo, S.R.S. Ferreira, D. Oliveira, Enzymatic ring opening copolymerization of globalide and ε-caprolactone under supercritical conditions, J. Supercrit. Fluids. 128 (2017) 404–411. doi:10.1016/j.supflu.2017.06.008.

[14] OECD, OECD 301 - Ready Biodegradability, OECD Guidel. Test. Chem. 301 (1992) 1–62. doi:10.1787/9789264070349-en.

[15] M.D. Hanwell, D.E. Curtis, D.C. Lonie, T. Vandermeersch, E. Zurek, G.R. Hutchison, Avogadro: an advanced semantic chemical editor, visualization, and analysis platform, J. Cheminform. 4 (2012) 17. doi:10.1186/1758-2946-4-17.

[16] J.X. Qin, M. Zhang, C. Zhang, C.T. Li, Y. Zhang, J. Song, H.M. Asif Javed, J.H. Qiu, New insight into the difference of PC lipase-catalyzed degradation on poly(butylene succinate)-based copolymers from molecular levels, RSC Adv. 6 (2016) 17896–17905. doi:10.1039/c5ra13738a.

[17] S.K. Saha, H. Tsuji, Hydrolytic degradation of amorphous films of L-lactide copolymers with glycolide and D-lactide, Macromol. Mater. Eng. 291 (2006) 357–368. doi:10.1002/mame.200500386.

[18] T. Casalini, Bioresorbability of polymers: chemistry, mechanisms, and modeling, in: Bioresorbable Polym. Biomed. Appl., Woodhead Publishing-Elsevier, Cambridge, 2017: pp. 65–83. doi:10.1016/B978-0-08-100262-9.00003-3.

[19] N.S. Allen, M. Edge, M. Mohammadian, M. Polytechnic, C. Street, M. Mi, K.E.N. Jones, Hydrolitic degradation of Poly (Ethylene Terephthalate): Importance of chain scission versus crystallinity, Eur. Polym. J. 27 (1991) 1373–1378.

[20] A. Vidaurre, J.M.M. Dueñas, J.M. Estellés, I.C. Cortázar, Influence of Enzymatic Degradation on Physical Properties of Poly ( e -caprolactone ) Films and Sponges, Macromol. Symp. 269 (2008) 38–46.

[21] G. Montaudo, P. Rizzarelli, Synthesis and enzymatic degradation of aliphatic copolyesters, Polym. Degrad. Stab. 70 (2000) 305–314. doi:10.1016/S0141-3910(00)00139-7.

[22] A.A. Gavrilov, Y. V. Kudryavtsev, P.G. Khalatur, A. V. Chertovich, Simulation of phase separation in melts of regular and random multiblock copolymers, Polym. Sci. Ser. A. 53 (2011) 827–836. doi:10.1134/s0965545x11090033.

[23] M. Vert, Aliphatic Polyesters: Great Degradable Polymers That Cannot Do Everything, Biomacromolecules. 6 (2005) 538–546. doi:10.1021/bm0494702.

[24] N.L. Davison, F.B. Groot, D.W. Grijpma, Degradation of Biomaterials, Second Edition, Elsevier Inc., 2015. doi:10.1016/B978-0-12-420145-3.00006-7.

[25] A. Göpferich, Mechanisms of polymer degradation and erosion, Biomaterials. 17 (1996) 103–114.

[26] C. V Paz, S.R. Vásquez, N. Flores, L. García, J.L. Rico, Reactive sites influence in

PMMA oligomers reactivity: a DFT study, Mater. Res. Express. 5 (2018) 15314. doi:10.1088/2053-1591/aaa679.

[27] N.M. Garrido, M. Jorge, A.J. Queimada, E.A. Macedo, I.G. Economou, Using molecular simulation to predict solute solvation and partition coefficients in solvents of different polarity, Phys. Chem. Chem. Phys. 13 (2011) 9155. doi:10.1039/c1cp20110g.

[28] S. Genheden, Solvation free energies and partition coefficients with the coarse-grained and hybrid all-atom/coarse-grained MARTINI models, J. Comput. Aided. Mol. Des. 31 (2017) 867–876. doi:10.1007/s10822-017-0059-9.

[29] J. Frau, D. Glossman-Mitnik, Conceptual DFT Study of the Local Chemical Reactivity of the Colored BISARG Melanoidin and Its Protonated Derivative, Front. Chem. 6 (2018). doi:10.3389/fchem.2018.00136.

[30] A. Rauk, Orbital Interaction Theory of Organic Chemistry, Second Edition., John Wiley & Sons, New York, 2001. doi:10.1002/aoc.182.

[31] J.M. Pachence, M.P. Bohrer, J. Kohn, Biodegradable polymers, in: R. Lanza, R. Langer, J. Vacanti (Eds.), Princ. Tissue Eng., 3rd Editio, Elsevier, 2007: pp. 323–339. doi:10.1016/B978-012370615-7/50027-5.

# TABLES

Table 1: Dipole moments, Gibbs free energy and $log\ P^{O/W}$ calculated at NPT using DFT/B3LYP/6-311++G** with water and n-octanol solvents in SMD model.

| Solvent | Parameter | PGlCL | PGlCL-NAC |
|---|---|---|---|
| Gas-Phase | $\mu$ (Debye) | 2.56 | 5.89 |
| | $G$ (Kcal/mol) | 313.935 | 400.031 |
| Water | $\mu_{solv}^{W}$ (Debye) | 3.77 | 8.39 |
| | $G_{solv}^{W}$ (Kcal/mol) | 313.934 | 400.028 |
| | $\Delta G_{solv}^{W}$ (Kcal/mol) | -7.66x10$^{-4}$ | -2.87x10$^{-3}$ |
| n-Octanol | $\mu_{solv}^{O}$ (Debye) | 3.52 | 7.88 |
| | $G_{solv}^{O}$ (Kcal/mol) | 306.690 | 395.512 |
| | $\Delta G_{solv}^{O}$ (Kcal/mol) | -7.24 | -4.52 |
| n-Octanol/Water | $log\ P^{O/W}$ | 5.31 | 3.31 |